\documentclass[a4paper,fleqn]{cas-sc}
\usepackage[numbers,sort&compress]{natbib}
\usepackage{caption} 

\usepackage{amsmath,amssymb,mathrsfs}
\usepackage{cases}
\usepackage{graphicx}% Include figure files
\usepackage{dcolumn}% Align table columns on decimal point
\usepackage{bm}% bold math
\usepackage{color, xcolor}
\usepackage{amsfonts}
\usepackage{enumerate}
\usepackage{rotating}
\usepackage{epstopdf}
\usepackage{braket}
\usepackage{bbding,pifont}

\hypersetup{colorlinks=true, citecolor=blue, linkcolor=blue, urlcolor=blue}

\usepackage{hyperref}
\hypersetup{
	colorlinks=true,
	linkcolor=blue,
	filecolor=blue,
	urlcolor=blue,
	citecolor=magenta
}

\newcommand{\Eq}[1]{Eq.~\eqref{#1}}
\newcommand{\eq}[1]{\eqref{#1}}
\newcommand{\Fig}[1]{Fig.~\ref{#1}}

\newcommand{\beq}{\begin{equation}}
	\newcommand{\eeq}{\end{equation}}
\newcommand{\beqa}{\begin{eqnarray}}
	\newcommand{\eeqa}{\end{eqnarray}}
\newcommand{\Beqa}{\begin{eqnarray*}}
	\newcommand{\Eeqa}{\end{eqnarray*}}

\def\bal#1\eal{\begin{align}#1\end{align}}
\def\Bal#1\Eal{\begin{align*}#1\end{align*}}

\newcommand{\nn}{\nonumber}

\newcommand{\me}{\mathrm{e}}

\newcommand{\mi}{\mathrm{i}}

\newcommand{\dif}{\mathrm{d}}

\newcommand{\vect}[1]{\mathbf{#1}}

\ExplSyntaxOn
\cs_gset:Npn \__first_footerline:
{ \group_begin: \small \sffamily \__short_authors: \group_end: }
\ExplSyntaxOff 
\begin{document}

% Short title
\shorttitle{Tailoring Synthetic Gauge Fields in Ultracold Atoms via Spatially Engineered Vector Beams}    

% Short author
\shortauthors{H. Wang et al.}

\let\WriteBookmarks\relax
\def\floatpagepagefraction{1}
\def\textpagefraction{.001}
\let\printorcid\relax
\let\footnote\gobble
	
\title[mode = title]{Tailoring Synthetic Gauge Fields in Ultracold Atoms via Spatially Engineered Vector Beams}

% 使用 \author 命令的正确格式
\author[1,2]{Huan Wang}\fnref{equal}
\credit{Methodology, Software, Validation, Formal analysis, Investigation, Writing - Original Draft, Visualization} 

\author[1,2,3]{Shangguo Zhu}\fnref{equal}
\credit{Conceptualization, Methodology, Validation, Formal analysis, Writing - Review \& Editing}  

\author[1,2,3]{Yun Long}
\credit{Conceptualization, Methodology, Writing - Review \& Editing} 

\author[1,2,3]{Mingbo Pu}
\credit{Supervision, Project administration, Funding acquisition} 

\author[1,3]{Xiangang Luo}\corref{cor1}
\credit{Supervision, Project administration, Funding acquisition} 
\ead{lxg@ioe.ac.cn}
\cortext[cor1]{Corresponding author}

\fntext[equal]{These authors contribute equally to this work.}

% 关联作者和单位(关键！)
\affiliation[1]{affiliation={State Key Laboratory of Optical Field Manipulation Science and Technology, Chinese Academy of Sciences, Chengdu 610209, China}}
\affiliation[2]{affiliation={Research Center on Vector Optical Fields, Institute of Optics and Electronics, Chinese Academy of Sciences, Chengdu 610209, China}}
\affiliation[3]{affiliation={College of Materials Science and Opto-Electronic Technology, University of Chinese Academy of Sciences, Beijing 100049, China}}

%%%%%%%%%%%%%%%%%%%%%%%%%%%%%%%%%%%%%  DATE  %%%%%%%%%%%%%%%%%%%%%%%%%%%%%%%%%%%%

%%%%%%%%%%%%%%%%%%%%%%%%%%%%%  Abstract  %%%%%%%%%%%%%%%%%%%%%%%%%%%%%
\begin{abstract}
Ultracold atoms, typically manipulated by scalar beams with uniform polarization, have propelled advances in quantum simulation, computation, and metrology. 
Yet, vector beams (VBs)---structured light with spatially varying polarization---remain unexplored in this context, despite their enhanced tunability and broad optical applications. 
Here, we demonstrate a novel scheme to generate synthetic gauge fields in ultracold atoms via VB-mediated coupling of internal states. 
This approach enables angular stripe phases across an expanded parameter range, achieving a three-order-of-magnitude enhancement in the phase diagram and facilitating experimental observation. 
We further present an all-optical method to create topologically nontrivial giant skyrmions in spin space, with tunable topology governed by VB parameters. 
Our findings establish VBs as powerful tools for quantum control and the exploration of exotic quantum states and phases. 
\end{abstract}


\maketitle

\section{\label{introduction} Introduction}

Ultracold atoms provide exceptional controllability, making them an ideal platform for quantum simulation~\cite{Bloch2012, Gross2017, Georgescu2014}, quantum computation~\cite{Saffman2010, Henriet2020}, and quantum metrology~\cite{Pezze2018}. 
A key direction is using ultracold atomic gases to simulate complex systems, including condensed matter models~\cite{Bloch2008}. 
A significant milestone in this endeavor is the creation of synthetic gauge fields via engineered optical coupling between atomic internal states~\cite{Lin2009, Lin2011a, Lin2011b, Goldman2014, Zhai2015}, particularly the realization of spin-orbit coupling (SOC)~\cite{Lin2011b, Ho2011, Wang2012, Cheuk2012, Zheng2013, Galitski2013, Goldman2014, Zhai2015, Huang2016, Wu2016, Wang2021a}, an essential ingredient in many condensed matter systems such as topological insulators, superconductors, and semimetals~\cite{Nagaosa2010, Qi2011, Armitage2018, Bihlmayer2022}. 
This progress has enabled the realization of numerous complex quantum phenomena~\cite{Kroeze2019, Putra2020, Li2019, Li2020, Liang2024, Mukhopadhyay2024, Liang2023, Pan2015, Wu2017, Wang2017, Cui2018, Deng2018, He2019, Wu2019, Tang2018, Zhu2021, Yasir2022}.

While synthetic SOC typically refers to the coupling between spins (atomic internal states, or pseudospins) and translational motion, spin-orbit-angular-momentum coupling (SOAMC)---a coupling between spins and rotational motion---has recently been proposed and realized in Bose-Einstein condensates (BECs)~\cite{DeMarco2015, Sun2015, Qu2015, Hu2015, Chen2016, Chen2020a}. 
In contrast to SOC which uses Gaussian beams, SOAMC employs Laguerre-Gaussian beams (LGBs) carrying finite orbital angular momenta (OAM).
Recent experiments have delineated ground‐state phase diagrams~\cite{Zhang2019} and vortex structure evolution~\cite{Chen2018} in SOAMC systems. 
Moreover, ultracold systems with SOAMC are theoretically predicted to host exotic quantum phases~\cite{Peng2022, Ng2023}, such as giant vortices in Fermi superfluids with vortex sizes comparable to the Raman beam waist~\cite{Chen2020b} and topological superfluids in ring‐shaped Fermi gases~\cite{Chen2022}.	
Most notably, the angular stripe phase in BECs~\cite{Chen2020c, Chen2020d, Duan2020, Chiu2020, Wang2021b}, which breaks $U(1)$ gauge symmetries and rotational symmetries and exhibits supersolid-like behavior in the rotational degree of freedom, remains experimentally elusive due to its narrow parameter window~\cite{Zhang2019, Chen2018, Qu2015}.
To broaden this window, it has been proposed to reduce the LGB waist using a high-numerical-aperture (NA) lens~\cite{Chen2018, Zhang2019}. 
Yet, tight focusing of conventional scalar beams with uniform polarization results in light fields lacking definite OAM required for SOAMC. 
A promising solution might lie in seeking novel coupling schemes using structured light fields~\cite{Forbes2021}, which provide greater flexibility in light field manipulation.

Vector beams, characterized by spatially varying polarization, offer extensive  tunable degrees of freedom, including polarization state, OAM, and beam shape~\cite{Rosales2018, Shen2019, Forbes2021}. 
These properties can be precisely engineered using spatial light modulators, phase plates and metasurfaces to create intricate vector light fields~\cite{Zhang2016, Rosales2018, Guo2021, Luo2022, Feng2023}.  
Such tailored light fields have enabled innovative applications in atomic physics~\cite{Wang2020}, such as three-dimensional magnetic field measurements~\cite{Castellucci2021}, and spatially dependent electromagnetically induced transparency~\cite{Radwell2015}. 
However, their potential for novel applications in ultracold atomic physics, quantum optics, and quantum information processing remains largely unexplored.

In this work, we introduce a novel scheme to couple atomic pseudospins in a BEC using two VBs through a tightly focusing system. 
This approach leverages the rich tunability of structured light to achieve enhanced quantum control in ultracold atomic systems. 	
Our setup generates a three-component synthetic magnetic field, contributing to SOAMC in the direction perpendicular to the applied magnetic filed $\vect{B}$ and a spatially dependent Zeeman shift along $\vect{B}$.  
The key innovation lies in achieving definite OAM transfer for SOAMC and tailoring the Zeeman shift’s spatial profile via precise VB engineering. 
This enables two significant demonstrations. 
First, considering SOAMC alone, we reveal an angular stripe phase with discrete rotational symmetry accessible over a significantly expanded experimental parameter range, achieving a \textit{three-orders-of-magnitude} enhancement in critical coupling strength compared to conventional LGB schemes~\cite{Zhang2019, Qu2015}. 
Second, by combining SOAMC with the spatially dependent Zeeman shift, we demonstrate an all-optical method for generating stable multiply quantized vortices---topologically nontrivial giant skyrmions~\cite{Yang2008, Mason2011, Jin2013, Dong2017} on the micrometer scale in spin space---with tunable topology via VB parameters. 
Finally, we discuss the realization of our scheme in current experiments.

\section{\label{coupling scheme} Coupling scheme}
\begin{figure}[t!]
	\centering
	\includegraphics[width = 0.5\linewidth]{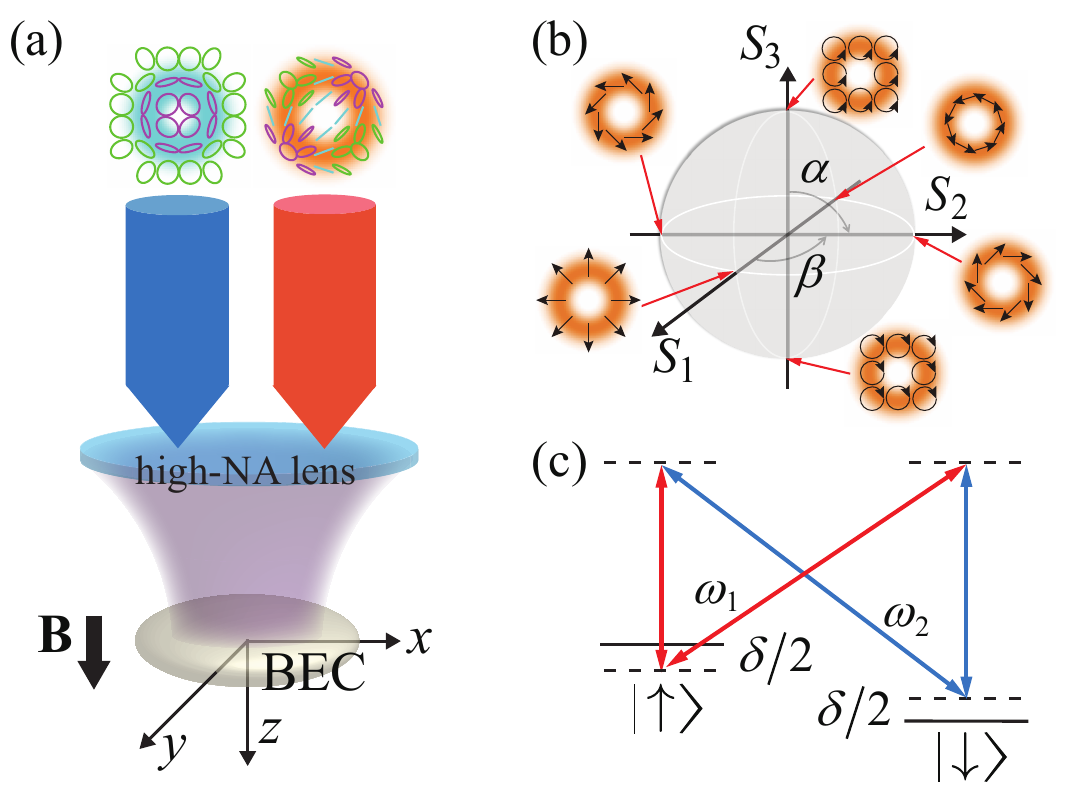}
	\caption{ 
		(a) Illustration of two VBs passing through a tightly focusing system and illuminating a pancake-shaped BEC trapped near the focal plane ($z=0$). 
		Insets at the top display the polarization and amplitude distributions at the focal plane for VBs with $m_1 = 1$, $l_1 = 2$, $\alpha_1  = \pi/2$,  $\beta_1 = 0$ (left) and $m_2 = 1$, $l_2 = 3$, $\alpha_2 = \pi/2$, $\beta_2 = 0$ (right) after focusing.  
		Magenta, green, cyan represent left-handed elliptical, right-handed elliptical, and linear polarizations, respectively. 
		(b) High-order Poincaré sphere for a VB (before focusing) with $m=1$ and $l=2$. 
		$S_1$, $S_2$, and $S_3$ are the Stokes parameters~\cite{Rosales2018}. 
		Insets show polarization (black arrows) and amplitude (orange backgrounds) distributions for different $\alpha$ and $\beta$. 
		(c) Atomic energy level structure with a double-$\Lambda$-type coupling scheme.  
	}
	\label{fig:Scheme} 
\end{figure}

The time-independent electric field of a VB can be expressed as 
\begin{align}
	\label{eq:E_incident}
	\boldsymbol{\mathcal{E}}(\rho,\varphi) = A_l(\rho, \varphi ) \Big[ \cos( \alpha/2 ) \me^{-\mi \beta/2}  \me^{-\mi m \varphi} \hat{\vect{e}}_L  +\sin( \alpha/2 ) \me^{\mi \beta/2} \me^{\mi m \varphi} \hat{\vect{e}}_R \Big], 
\end{align}
where $(\rho, \varphi, z)$ are cylindrical coordinates, with $\rho$ and $\varphi$ denoting the radius and azimuth angle, respectively, and the light propagates along the $z$-axis. 
The unit vectors $\hat{\vect{e}}_{L,R} \equiv (\hat{\vect{e}}_x \pm \mi \hat{\vect{e}}_y)/\sqrt
2$ correspond to left (L) and right (R) circular polarization states. 
The amplitude $A_l(\rho, \varphi )  = \sqrt{I} ( \sqrt{2}\rho /w )^{|l|} \me^{-\rho^2/w^2 }  \me^{\mi l\varphi}$ describes an LGB profile, where $w$ is the beam waist, $I$ is the light intensity, and $l$ is the vortex topological charge.	
The combination of $\me^{-\mi m \varphi} \hat{\vect{e}}_L$ and $\me^{\mi m \varphi} \hat{\vect{e}}_R$ can be described by the high-order Poincaré sphere~\cite{Rosales2018}, with $(\alpha, \beta)$ as spherical coordinates ($0\le \alpha \le \pi, 0\le \beta \le 2\pi$), as illustrated in \Fig{fig:Scheme} (b). 
The parameter $m$ denotes the polarization topological charge. 

Consider two VBs, labeled $j=1, 2$, with optical frequencies $\omega_j$ and electric fields $\boldsymbol{\mathcal{E}}_j$ described by \Eq{eq:E_incident} with parameters $(m_j, l_j, \alpha_j, \beta_j)$. 
For simplicity, both VBs are assumed to have the same waist $w$. 
After passing through a high-NA lens, they form tightly focused VBs~\cite{Chen2012, Yu2020}, illuminating a pancake-shaped BEC near the focal plane, as shown in \Fig{fig:Scheme} (a).
The tightly focused VBs exhibit three nonzero spatial components of their electric fields, $\vect{E}_j = (E_{xj}, E_{yj}, E_{zj})$ (see~\ref{app:electricfields} for explicit forms).  
These components are tunable via $(m_j, l_j, \alpha_j, \beta_j)$, offering great experimental flexibility to generate various light fields~\cite{Zhang2016, Rosales2018, Guo2021, Luo2022, Feng2023}. 
An external Zeeman field $\vect{B}$ is applied along the $z$-axis, inducing an energy splitting $\hbar \omega_{\mathrm{Z}}$ between the two spin states $\left|  \uparrow  \right\rangle$ and $\left|  \downarrow  \right\rangle$, where $\hbar = h/2\pi$ is the reduced Planck constant. 
As shown in \Fig{fig:Scheme} (c), $\left|  \uparrow  \right\rangle$ and $\left|  \downarrow  \right\rangle$ are coupled through two-photon Raman transitions with detuning $\delta= \hbar (\omega_{\mathrm{Z}} + \omega_1 - \omega_2 )$, forming a double-$\Lambda$-type coupling scheme (see~\ref{app:singleparticle} for details).

When interatomic interactions are weak, the motion of atoms can be approximated as a two-dimensional problem~\cite{Chen2020b, Zhang2019, Chen2018, Qu2015}. 
Here we focus on the physics at the focal plane. 
The system's dynamics and equilibrium properties are effectively governed by the Gross-Pitaevskii (GP) equation $\mi \hbar \partial_t \Psi = H \Psi$, where $\Psi = ( \psi_{\uparrow}, \psi_{\downarrow} )^{T}$ is the spinor wave function of the BEC. 
The GP Hamiltonian (see~\ref{app:singleparticle}) is 
\begin{equation}
	H=-\frac{\hbar^{2}\nabla_\rho^{2}}{2M}+\frac{1}{2}M \omega^{2} \rho^{2}+V_{\rm S} (\boldsymbol{\rho})+V_{\mathrm{VB}}(\boldsymbol{\rho})+V_{\mathrm{I}}(\boldsymbol{\rho})+\frac{\delta}{2}\sigma_{z}, \label{eq:HGP}
\end{equation}
where $\boldsymbol{\rho} \equiv (x, y)$ is the position in the $x-y$ plane, $M$ is the atomic mass, $\omega$ is the transverse trapping frequency (with trap size $a_0 \equiv \sqrt{\hbar/M\omega}$), and $\vec{\sigma}_{\rm P} \equiv (\sigma_x, \sigma_y, \sigma_z)$ represents the Pauli matrices. 
$V_{\rm S}(\boldsymbol{\rho})$, $V_{\mathrm{VB}}(\boldsymbol{\rho})$, and $V_{\mathrm{I}}(\boldsymbol{\rho})$ are the potential induced by scalar light shift, the potential associated with the effective magnetic field induced by vector light shift, and the nonlinear mean-field interaction, respectively: 
\begin{align}
	V_{\rm S} (\boldsymbol{\rho}) &= \Omega_{s}(\bar{\mathbf{E}}_{1}^{*} \cdot \bar{\mathbf{E}}_{1}+\bar{\mathbf{E}}_{2}^{*} \cdot \bar{\mathbf{E}}_{2}), \label{eq:Vsr} \\
	V_{\mathrm{VB}}(\boldsymbol{\rho}) &= \begin{pmatrix} \Omega_{z}(\boldsymbol{\rho}) & \Omega_{r}(\boldsymbol{\rho}) \\ \Omega_{r}^{*}(\boldsymbol{\rho}) & -\Omega_{z}(\boldsymbol{\rho})\end{pmatrix}, \label{eq:VVB}  \\
	V_{\mathrm{I}}(\boldsymbol{\rho}) &= \eta \begin{pmatrix} g_{\uparrow\uparrow}|\psi_{\uparrow}|^2+g_{\uparrow\downarrow}|\psi_{\downarrow}|^2 & 0 \\ 0 & g_{\uparrow\downarrow}|\psi_{\uparrow}|^2+g_{\downarrow\downarrow}|\psi_{\downarrow}|^2 \end{pmatrix}. \label{eq:VI}
\end{align}
Here $\bar{\mathbf{E}}_{j} \equiv \mathbf{E}_{j}/\sqrt{I_j} \equiv (\bar{E}_{xj}, \bar{E}_{yj}, \bar{E}_{zj})$ with $I_j$ the light intensity.  
The off-diagonal term of $V_{\mathrm{VB}}$, 	$\Omega_{r}(\boldsymbol{\rho}) = \Omega_{0} [(\bar{E}_{z2}^{*}\bar{E}_{x1} - \bar{E}_{x2}^{*}\bar{E}_{z1}) + \mi (\bar{E}_{y2}^{*}\bar{E}_{z1} - \bar{E}_{z2}^{*}\bar{E}_{y1}) ]$, drives transitions between Zeeman sublevels, coupling spin and OAM. 
The diagonal term of $V_{\mathrm{VB}}$, $\Omega_{z}(\boldsymbol{\rho}) = \mi \Omega_{0}  (\bar{E}_{x1}^{*}\bar{E}_{y1} - \bar{E}_{y1}^{*}\bar{E}_{x1} + \bar{E}_{x2}^{*}\bar{E}_{y2} - \bar{E}_{y2}^{*}\bar{E}_{x2})$, absent in LGB-induced SOAMC schemes~\cite{DeMarco2015, Sun2015, Qu2015, Hu2015, Chen2016, Chen2020a}, introduces a spatially dependent Zeeman shift, providing additional control. 
The interaction strength $g_{\sigma\sigma^{\prime}} = 4\pi\hbar^2Na_{\sigma\sigma^{\prime}}/M$, where $a_{\sigma\sigma^{\prime}}$ is the $s$-wave scattering length between spins $\sigma = \uparrow, \downarrow$, and $N$ is the atom number. 
The parameter $\eta = 1/(\sqrt{2\pi} a_{0z})$ is a dimensional reduction factor, with $a_{0z}$ representing the atom cloud width along the $z$-axis (see~\ref{app:interactionBEC}).

\section{\label{angularstripephase}Angular stripe phase }
The multiple tunable degrees of freedom in VBs allow for various outcomes based on the Hamiltonian $H$ in \Eq{eq:HGP} for different VB combinations (see~\ref{app:Potential}). 
For simplicity, we consider the case $m_1 = m_2 = 1$ and $l_1 = - l_2 = n$, where $n$ is a tunable integer. 
Without loss of generality, we set $\alpha_1 = \alpha_2 = \pi/2$ and $\beta_1 = \beta_2 = 0$. 
Under this configuration, $V_{\mathrm{VB}} (\boldsymbol{\rho})$ contains only the off-diagonal term, with elements given by (see~\ref{app:Potential} Table~\ref{tab:varyingVBParameters}) 
\begin{equation}
	\Omega_z=0, \qquad \Omega_{r}=\Omega_{0} f(\rho) \me^{\mi \Delta l \varphi},  \label{eq:omegar}
\end{equation}
where $f(\rho)$ describes the spatial distribution and $\Delta l = l_1 - l_2 - 1 = 2n-1$ represents the OAM transfer during the Raman process.

\begin{figure}[t!]
	\centering
	\includegraphics[width = 0.5\linewidth]{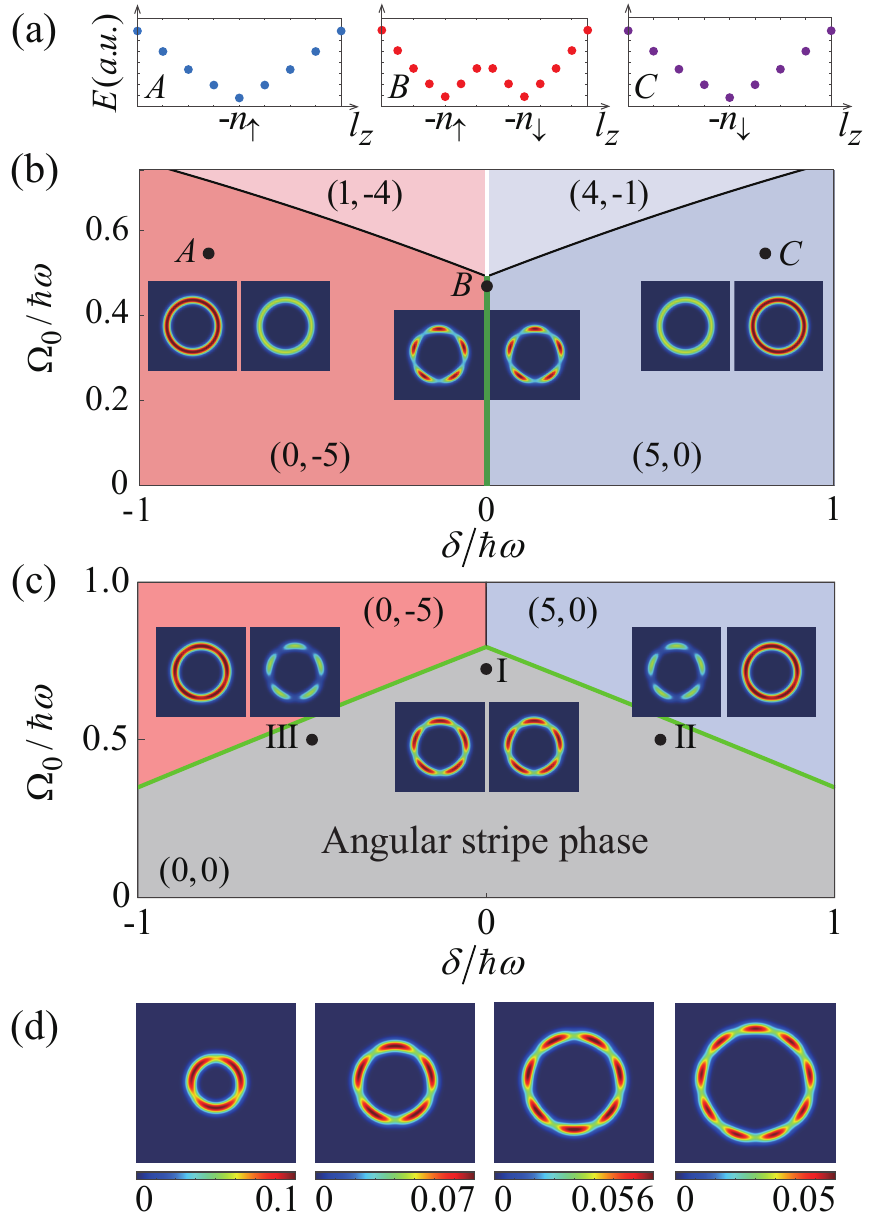} 
	\caption{ 
		(a) Single-particle energy spectra corresponding to points $A$ (left), $B$ (middle), and $C$ (right) in (b). 
		(b) Single-particle phase diagram in the $\Omega_0-\delta$ plane for $n=3$ and $\Omega_s/\hbar\omega=-13$. 
		Phases are labeled by $(m_{\uparrow}, m_{\downarrow})$. 
		The green solid line represents the angular stripe phase for $\delta = 0$ and $0\le \Omega_0 \le \Omega_0^c $, where $\Omega_0^c\approx 0.48 \hbar \omega$. 
		Insets show normalized density distributions $|\psi_\sigma|^2 a_0^2 /N$ (left for $\sigma=\uparrow$, right for $\sigma=\downarrow$) at points $A$, $B$, and $C$, with $(\delta, \Omega_{0})/\hbar \omega = (-0.8, 0.55)$, $(0, 0.47)$, and $(0.8, 0.55)$, respectively. 
		The spin $\downarrow$ density distribution at $A$ and $\uparrow$ at $C$ are magnified $10$ times for viewing. 
		(c) Phase diagram for weakly interacting BECs with scattering lengths $a_{\uparrow \uparrow} = a_{\downarrow \downarrow} = 45 a_B$ and $a_{\uparrow \downarrow} = 0.5 a_{\uparrow \uparrow}$, where $a_B$ is the Bohr radius. 
		Other parameters are the same as in (b). 
		The angular stripe phase (gray) achieves a critical coupling strength $\Omega_0^c \approx 0.82\hbar \omega$ at $\delta=0$. 
		Insets show normalized density distributions of the angular stripe phase at points I, II, and III, with $(\delta, \Omega_{0})/\hbar \omega = (0, 0.75)$, $(0.5, 0.5)$, and $(-0.5, 0.5)$, respectively. 
		(d) Angular stripe phases with varying periodicity.  $|\psi_\uparrow|^2 a_0^2 /N$ are shown for $x,y \in [-5 a_0, 5 a_0]$, at $\delta = 0$ and the following parameters: 
		(left) $n=2$, $\Omega_s/\hbar \omega = -29.27$, $\Omega_0/\hbar\omega = 1.63$;
		(middle left) $n=3$, $\Omega_s/\hbar \omega = -13$, $\Omega_0/\hbar\omega = 0.75$ (same as point I in (c)); 
		(middle right) $n=4$, $\Omega_s/\hbar \omega = -3.9$, $\Omega_0/\hbar\omega = 0.2$; 
		(right) $n=5$, $\Omega_s/\hbar \omega = -0.85$,  $\Omega_0/\hbar\omega = 0.04$. 
	}
	\label{fig:PhaseDiagram} 
\end{figure}

We first focus on single-particle physics, neglecting interatomic interactions. 
Defining two OAMs $n_{\uparrow} \equiv n-1$ and $n_{\downarrow} \equiv -n$, we transform the basis states as $\psi_{\sigma} \to \me^{\mi  n_{\sigma} \varphi} \psi_{\sigma}$. 
Substituting into \Eq{eq:HGP}, the single-particle Hamiltonian becomes
\begin{align}
	H_{\mathrm{0}}&=-\frac{\hbar^{2}}{2M}\frac{1}{\rho}\frac{\partial}{\partial \rho} \left( \rho\frac{\partial}{\partial \rho } \right)+\frac{L_{z}^{2}}{2M \rho^{2}} + \frac{1}{2}M\omega^{2} \rho^{2}+V_{\rm S}(\rho)  + \frac{\hbar\gamma}{M \rho^{2}}L_{z} +\frac{\hbar^{2}\gamma^{2}}{2M \rho^{2}} +\frac{\delta}{2}\sigma_{z} + \Omega_{r}(\rho)\sigma_{x},  \label{eq:H0_gamma} 
\end{align}
where $L_z=-\mi \hbar\partial_\varphi$ is the quasiangular momentum (QAM) operator, and $\gamma = \frac{n_{\uparrow}+n_\downarrow}{2} +\frac{n_{\uparrow}-n_{\downarrow}}{2} \sigma_z$ is a scaling matrix. 
Since $[L_z, H_0] = 0$, each eigenstate has a definite QAM $l_{z}$, which is related to the OAM $m_\sigma$ of spin component $\sigma$ in the laboratory frame via $m_\sigma=l_{z}+n_\sigma$. 
The term $\hbar\gamma L_{z} / M \rho^2$ introduces coupling between spin and OAM when $n_{\uparrow} \neq n_{\downarrow}$.  
Due to rotational symmetry, the wave function can be expressed as $\psi_{\sigma} (\boldsymbol{\rho}) = \bar{f}_{\sigma \tilde{n}}(\rho) \me^{\mi l_z \varphi}/\sqrt{2\pi}$, where $\tilde{n}$ is the radial quantum number. 
We solve $H_{\mathrm{0}} \Psi = E \Psi$ to obtain the single-particle energy spectrum~\cite{Chen2020b, Chen2020d} (\Fig{fig:PhaseDiagram} (a)). 
By varying $\Omega_0$ and $\delta$, we compute the single-particle phase diagram (\Fig{fig:PhaseDiagram} (b)) using imaginary time evolution.

At zero detuning $\delta$ and small coupling strength $\Omega_0$, the ground state typically exhibits twofold degeneracy at $l_{z} = -n_{\uparrow}$ and $l_z = -n_{\downarrow}$, as seen in the middle panel of \Fig{fig:PhaseDiagram} (a). 
The density modulation can be described by~\cite{Chiu2020}
\begin{equation}
	|\psi_{\uparrow}|^{2} = |\psi_{\downarrow}|^{2} = \bar{n}_0(\rho) \left\{ \frac{1}{2} + \frac{1}{2} \bar{n}_1(\rho) \cos[ \Delta l \varphi + \bar{\varphi}] \right\}, 
\end{equation}
where $\bar{n}_0(\rho)$ is the azimuthally averaged density, satisfying $\int \dif\rho~ 2\pi \rho \bar{n}_0 (\rho) = N$, $\bar{n}_1(\rho)$ is the modulation contrast, and $\bar{\varphi}$ is a phase constant. 
This describes an azimuthal density modulation with a period of $2\pi/\Delta l$, forming an angular stripe phase. 
Consistent with our numerical results, the angular stripe phase exhibits a discrete $(2n-1)$-fold rotational symmetry, as shown in the inset (point B) of \Fig{fig:PhaseDiagram} (b).

For nonzero detuning $\delta$, the ground-state degeneracy is lifted, and the ground-state energy localizes at a single minimum, as shown in the left and right panels of \Fig{fig:PhaseDiagram} (a). 
At $\delta<0$, increasing $\Omega_0$ from $0$, the phase diagram splits into regions with ground-state QAM $l_z = -n_{\uparrow}, -n_{\uparrow}+1, \cdots$, corresponding to OAM $(m_\uparrow, m_\downarrow) = (0, -2n+1), (1, -2n+2), \cdots$. 
At $\delta>0$, increasing $\Omega_0$ from $0$, the phase diagram splits into regions with $l_z = -n_{\downarrow}, -n_{\downarrow}-1, \cdots$, corresponding to $(m_\uparrow, m_\downarrow) = (2n-1, 0), (2n-2, -1), \cdots$. 
These ground states correspond to spin-polarized phases without azimuthal density modulation, as illustrated in the insets (points A and C) of \Fig{fig:PhaseDiagram} (b).

Now we examine the phase diagram with weak interatomic interactions, focusing on the case $g_{\uparrow \uparrow} > g_{\uparrow \downarrow}$, which favors a balanced density profile between spin components.
Using the imaginary time evolution method, we obtain the ground-state phase diagram shown in \Fig{fig:PhaseDiagram} (c). 
Compared to the single-particle case, interactions alter the phase boundaries. 
The angular stripe phase with $(m_\uparrow, m_\downarrow) = (0, 0)$ expands from a line into a broader region, as interactions serve as a stabilizing factor against detuning, thereby extending its parameter range. 
For $\delta=0$, the system exhibits a spin-balanced angular stripe phase, while for $\delta \neq 0$, it transitions to a spin-imbalanced angular stripe phase, absent in the single-particle scenario. 
Increasing the coupling strength $\Omega_0$ beyond a critical value $\Omega_0^c$ leads to spin-polarized phases, with $(m_\uparrow, m_\downarrow) = (0, -2n+1)$ for $\delta<0$, or $(m_\uparrow, m_\downarrow) = (2n-1, 0)$ for $\delta>0$, both without azimuthal density modulation.

Enabled by the rich tunability of VBs, the angular stripe phase exhibits tunable $C_{\Delta l}$ discrete rotational symmetry, where $\Delta l = 2n-1$ and $n$ is controlled by VB parameters. 
By varying $n$, angular stripe phases with arbitrary odd periodicities can be achieved.
As shown in \Fig{fig:PhaseDiagram} (d), $C_{3}$, $C_{5}$, $C_{7}$, and $C_{9}$ rotational symmetries can be achieved by selecting appropriate values of $n$, $\Omega_s$, and $\Omega_0$, while keeping other parameters consistent with \Fig{fig:PhaseDiagram} (c).
As $n$ increases, the ring size of the angular stripe phase grows, while the required coupling strength $\Omega_0$ decreases, indicating a shrinking phase diagram region and a narrower experimental parameter window. 
Therefore, smaller $n$ values would be more experimentally feasible.

\section{\label{spintexture}Spin texture}
Our coupling scheme with VBs can generate exotic spin textures. 
The potential $V_{\mathrm{VB}} (\boldsymbol{\rho})$ can include both nonzero $\Omega_r$ and $\Omega_z$ in \Eq{eq:VVB}, whereas $\Omega_z$ is absent in LGB-induced SOAMC. 
For parameters $\beta_1 = \beta_2 = 0$, $m_1 = m_2 = 1$, $l_1 = 2$, and $l_2 = -4$, the elements of 
$V_{\mathrm{VB}} (\boldsymbol{\rho})$ become (see~\ref{app:Potential} Table~\ref{tab:varyingVBParameters})
\begin{equation}
	\Omega_z = \Omega_{0} f_z(\rho), \qquad \Omega_r = \Omega_{0} f(\rho) \me^{\mi \Delta l \varphi}, 
\end{equation}
where $f_z(\rho)$ and $f(\rho)$ describe the spatial distributions, and the OAM transfer $\Delta l = l_1 - l_2 - 1 = 5$.  
For simplicity, we set $\Omega_s=0$, achievable by choosing the tune-out wavelength. 
The GP equation with vanishing interaction is solved for various $\alpha_1$ and $\alpha_2$ to analyze the ground-state spin textures.

\begin{figure}[t!]
	\centering
	\includegraphics[width = 0.5\linewidth]{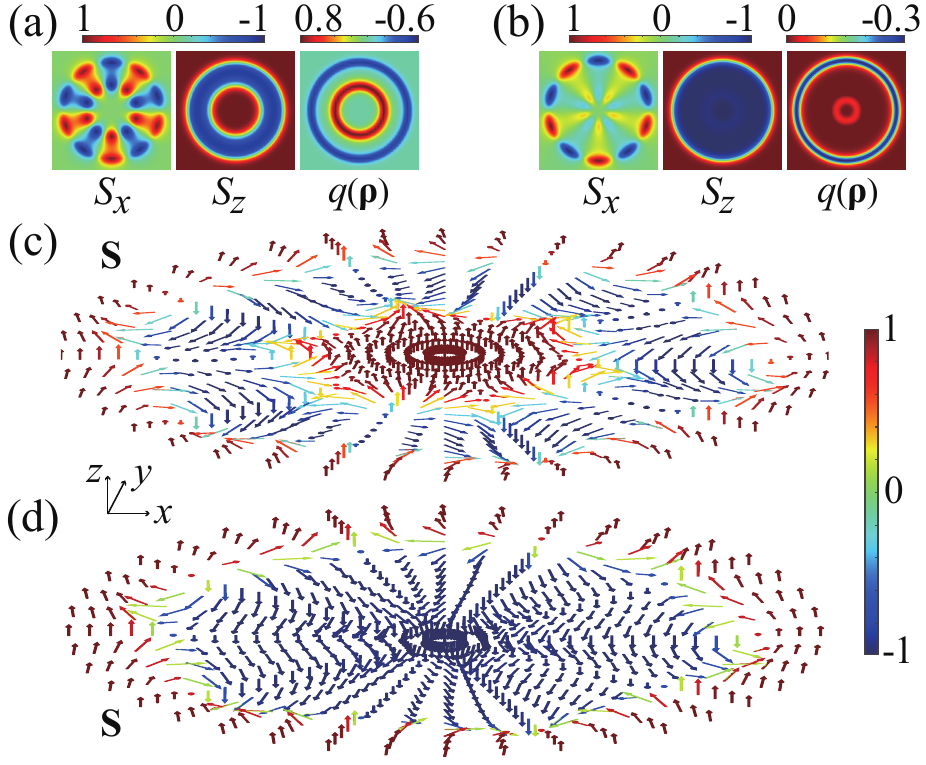}
	\caption{
		Spin textures and topological charge densities. 
		(a) and (c) correspond to $\alpha_1 = \alpha_2 = 0.37\pi$, while (b) and (d) correspond to $\alpha_1 = \alpha_2 = 0.45\pi$. 
		In (a) and (b), the left, middle, and right panels display $S_x$, $S_z$, and $q(\boldsymbol{\rho})$, respectively, with spatial ranges $x,y \in [-3.5 a_0, 3.5 a_0]$ in (a) and $x,y \in[-5 a_0, 5 a_0]$ in (b). 
		In (c) and (d), arrows indicate the distribution of $\vect{S}$, and the color bar represents $S_z$. }
	\label{fig:Spintexture}  
\end{figure}

The ground-state spin texture is described by the spin density vector $\vect{S}=\Psi^{\dagger} \vec{\sigma}_{\rm P} \Psi/|\Psi|^2$. 
For $\alpha_1=\alpha_2=0.37\pi$, and $\alpha_1=\alpha_2=0.45\pi$, $S_x$, $S_z$, and $\vect{S}$ are shown in \Fig{fig:Spintexture} (a) (c) and (b) (d), respectively (see~\ref{app:Skyrmion} for details on density profiles and relative phases). 
Both $S_x$ and $S_y$ exhibit periodic modulation in the azimuthal direction, with a periodicity determined by $\Delta l$, similar to the angular stripe phase.

The ground state exhibits a stable multiply quantized vortex with its quantized circulation determined by $\Delta l$ (see~\ref{app:Skyrmion}). 
For $\alpha_1 = \alpha_2 = 0.37\pi$, it forms two topological giant skyrmions in spin space~\cite{Yang2008, Mason2011, Jin2013, Dong2017}, while for $\alpha_1 = \alpha_2 = 0.45\pi$, there is one. 
Along the radial direction, $\vect{S}$ flips from north to south (or vice versa) when crossing the annular giant skyrmions. 
These topological structures are characterized by the topological charge density $q(\boldsymbol{\rho})$ (right panels of \Fig{fig:Spintexture} (a) and (b)) and the topological charge $Q$~\cite{Mason2011, DeMarco2015, Yang2008, Jin2013, Wang2017, Li2020}, defined as
\begin{equation}
	q(\boldsymbol{\rho}) = \vect{S} \cdot (\partial_x \vect{S} \times \partial_y \vect{S})/4\pi, \qquad
	Q  = \int_{\mathcal{A}} \dif^2 \rho~ q(\boldsymbol{\rho}), 
\end{equation} 
where $\mathcal{A}$ is the annular region $\rho_1 < \rho < \rho_2$ enclosing the skyrmion of interest. 
At $\alpha_1=\alpha_2=0.37\pi$, the inner ring has topological charge $Q_\mathrm{in}=5$ and the outer ring $Q_\mathrm{out}=-5$. 
Increasing $\alpha_1$ and $\alpha_2$ to $\alpha_1=\alpha_2=0.45\pi$ alters the spatial distribution of $\Omega_z$, reversing the spin imbalance in the inner region (see~\ref{app:Skyrmion}), and resulting in a single skyrmion with topological charge $Q=-5$. 
The absolute values of the topological charges are determined by $\Delta l$. 
Previously, generating such giant skyrmions required adding rotations~\cite{Yang2008, Mason2011, Jin2013, Dong2017}. 
In contrast, our scheme provides a novel all-optical approach to create giant skyrmions without rotation, and with their topology tunable via VB parameters.

\section{\label{experimentobservation}Experiment observation}
In conventional LGB-induced SOAMC systems, the angular stripe phase occupies a tiny region in the phase diagram, with a critical coupling strength $\Omega_0^c \approx h \times 0.0155 \mathrm{Hz}$~\cite{Qu2015, Zhang2019} at $\delta = 0$, which is nearly unattainable experimentally~\cite{Zhang2019, Chen2018}. 
While reducing the LGB waist using a high-NA lens has been proposed to expand this region~\cite{Chen2018, Zhang2019}, focusing LGBs introduces extra terms in $\Omega_{r}$ (see~\ref{app:Potential}), which lack a definite OAM transfer as in \Eq{eq:omegar} and cannot be eliminated due to the limited tunability of LGBs, thus preventing the emergence of SOAMC. 
In contrast, our VB-based scheme enables SOAMC by appropriately tuning VB parameters.

Consider $N=1000$ ${}^{87}$Rb atoms in an optical dipole trap with trapping frequencies $\omega_x = \omega_y = \omega = 2\pi \times 103.3 \mathrm{Hz}$ and $\omega_z = 10 \omega$, and transverse trap size $a_0 \approx 1.06 \mathrm{\mu m}$. 
We can choose the energy levels as the Zeeman sublevels $\left| \downarrow \right\rangle \equiv \left| 5^2S_{1/2}, F=1, m_F = 0 \right\rangle$ and $\left| \uparrow \right\rangle \equiv  \left| 5^2S_{1/2}, F=1, m_F=-1 \right\rangle$, and the excited states  $|5^2P_{1/2}, F=1, m_F = 0 \rangle$ and $|5^2P_{1/2}, F=1, m_F = -1 \rangle$, forming a double-$\Lambda$-type configuration as shown in \Fig{fig:Scheme} (c) (also see~\ref{app:singleparticle}). 	
The states $\left| \uparrow \right\rangle$ and $\left| \downarrow \right\rangle$ are coupled via two-photon Raman processes, obeying the corresponding selection rules.

Two characteristic energy scales are relevant~\cite{Zhang2019, Chen2018, Qu2015}: $E_r = \hbar^2 / 2 M \bar{w}^2$, characterizing the energy transferred during the Raman process, and $E_L = \Delta l^2 / 2MR^2$, characterizing the rotational energy with OAM $\Delta l$, where $\bar{w}$ is the spot size of tightly focused VB, and $R$ is the atom cloud radius~\cite{Zhang2019}. 
For the setup in \Fig{fig:PhaseDiagram} (c), with $\Delta l = 5$ and $R = 3 \mathrm{\mu m}$, we find $E_L \approx h \times 161.35 \mathrm{Hz}$. 
The tightly focused VBs form a doughnut-shaped spot with peak intensity at radius $r_M \approx 2.34 a_0 \approx  2.48 \mathrm{\mu m}$.
Setting $\bar{w} = r_M $, we obtain $E_r \approx h \times 9.43 \mathrm{Hz}$, giving $E_L / E_r \approx 17$. 
The critical coupling strength is $\Omega_0^c \approx 0.82\hbar \omega \approx h \times 84.71 \mathrm{Hz}$ at $\delta = 0$, which creates a significantly expanded angular stripe phase region, making it readily achievable in experiments. 
Thus, compared to LGB-induced SOAMC, our VB-based scheme provides a three-orders-of-magnitude enhancement, enabling feasible experimental observation.

\section{\label{sec4}Summary}
We have introduced a novel scheme employing VBs to couple the internal states of ultracold atoms, enabling tailored synthetic gauge fields by leveraging the exceptional tunability of VBs. 
With SOAMC alone, the ground-state phase diagrams reveal a significantly enhanced angular stripe phase, characterized by azimuthal modulation with discrete rotational symmetry and tunable periodicity via VB parameters. 
Compared to conventional LGB schemes, our approach expands the accessible phase region by three orders of magnitude, making experimental observation feasible.

Moreover, the angular stripe phase can be regarded as a precursor to a supersolid state. 
Its rotational symmetry makes properties such as the non-classical moment of inertia and superfluid fraction more tractable, establishing them as effective indicators of supersolidity~\cite{Cooper2010, Tanzi2021}. 
This paves the way for exploring supersolids with azimuthal modulation.

Furthermore, by incorporating both SOAMC and the spatially dependent Zeeman shift, we have demonstrated a mechanism for generating topologically nontrivial giant skyrmions without requiring rotation. 
This allows precise control of topology through all-optical methods, which could advance the study of skyrmion physics~\cite{Fert2013, Psaroudaki2021}.  
Beyond bosonic systems, our framework can also be extended to Fermi gases and optical lattices, providing a versatile toolbox for quantum control and the study of exotic quantum phenomena.

\printcredits

\section*{Declaration of competing interest}
The authors declare that they have no known competing financial interests or personal relationships that could have appeared to influence the work reported in this paper. 

\section*{Data availability}
Data will be made available on request.

\section*{Acknowledgements}
We acknowledge helpful discussions with Jiansong Pan, Shanshan Ding, Tianyou Gao, and Lingran Kong. 
S.Z. acknowledges support from the Youth Innovation Promotion Association, Chinese Academy of Sciences (2023399). 
X.L. acknowledges support from the National Natural Science Foundation of China (U24A6010, 52488301).

\renewcommand{\thesection}{A-\arabic{section}}
\setcounter{section}{0}  %  this will re-count section from 1
\renewcommand{\theequation}{A\arabic{equation}}
\setcounter{equation}{0}  %  this will re-count eq from 1
\renewcommand{\thefigure}{A\arabic{figure}}
\setcounter{figure}{0}  %  this will re-count eq from 1

\appendix
%\numberwithin{equation}{section}

% 重定义附录标题格式
\makeatletter
\renewcommand{\thesection}{Appendix \Alph{section}}
\makeatother

\section{\label{app:electricfields} Electric fields of tightly focused VBs}
We introduce the electric fields of tightly focused VBs after passing through a high-NA lens, as shown in Fig.~\ref{fig:Scheme}.  
The electric field at the focal plane are determined using Richards-Wolf vectorial diffraction integral~\cite{Richards1959, Chen2012, Yu2020} in the cylindrical coordinates $ (\rho, \varphi, z)$, given by:  
\begin{align}
	E_x & = \mi^{a+1} \cos(\alpha/2) \me^{-\mi \beta/2}
	\Big[\me^{\mi a\varphi}\mathcal{I}_a + \me^{\mi(a+2)\varphi}\mathcal{I}''_{a+2} \Big] + \mi^{b+1} \sin(\alpha/2) \me^{\mi \beta/2} \Big[\me^{\mi b\varphi}\mathcal{I}_b + \me^{\mi(b-2)\varphi}\mathcal{I}''_{b-2} \Big],\label{eq:Ex} \\
	E_{y} & = \mi^{a+2} \cos(\alpha/2)\me^{-\mi \beta/2} \Big[\me^{\mi a\varphi}\mathcal{I}_a-\me^{\mi(a+2)\varphi}\mathcal{I}''_{a+2} \Big]  + \mi^b \sin(\alpha/2) \me^{\mi \beta/2} \Big[\me^{\mi b\varphi}\mathcal{I}_b-\me^{\mi(b-2)\varphi}\mathcal{I}''_{b-2} \Big],\label{eq:Ey} \\
	E_z & = -\mi^{a+2} \cos (\alpha/2) \me^{-\mi \beta/2}  \me^{\mi(a+1)\varphi}\mathcal{I}'_{a+1} - \mi^b \sin(\alpha/2) \me^{\mi \beta/2} \me^{\mi (b-1)\varphi}\mathcal{I}'_{b-1}, \label{eq:Ez}
\end{align}
where $a  \equiv l-m$, and $b \equiv l+m$. 
$\alpha$ and $\beta$ are spherical coordinates of high-order Poincaré sphere as shown in \Fig{fig:Scheme}~(b).  
$\mathcal{I}_n$, $\mathcal{I}'_n$, and $\mathcal{I}''_n$ depend on spatial coordinates $\rho$ and $z$, expressed as
\begin{align}
	\mathcal{I}_n &= -kF\int_0^{\theta_{\max}}\dif \theta~  \bar{A}_l(\theta)\sin\theta \cos^{1/2}(\theta) (1+\cos\theta) e^{\mi kz\cos\theta}J_n(k \rho \sin\theta)  ,  \\
	\mathcal{I}'_n &=
	-2kF\int_0^{\theta_{\max}}\dif \theta~  \bar{A}_l(\theta)\sin^2\theta \cos^{1/2}(\theta)  e^{\mi kz\cos\theta}J_n(k \rho \sin\theta) ,    \\
	\mathcal{I}''_n &= -kF\int_0^{\theta_{\max}} \dif \theta~ \bar{A}_l(\theta)\sin\theta \cos^{1/2}(\theta) (1-\cos\theta)  e^{\mi kz\cos\theta}J_n(k \rho \sin\theta),  \label{eq:I_all}
\end{align}
where $k=2\pi/\lambda$ is the the wave number, $\lambda$ is the wavelength, and $J_n(x)$ denotes the $n$-th order Bessel function of the first kind. 	
The angle $\theta$ is defined for the point $(\rho, \varphi, z=-F)$ on the plane where the high-NA lens (with focal length $F$) is located, such that $\sin \theta = \rho/ F$ and  $0 \le \theta \le \theta_{\text{max}}$ with $\sin\theta_{\text{max}} = \mathrm{NA}$. 
The amplitude profile $\bar{A}_l(\theta)$ is given by
\begin{equation}
	\bar{A}_l (\theta) = \sqrt{I}\left[ \frac{\sqrt{2}\zeta\sin(\theta)}{\sin(\theta_{\max})} \right]^{|l|}\exp \left[ -\zeta^2 \left(\frac{\sin(\theta)}{\sin(\theta_{\max})}\right)^2 \right], \label{eq:A_theta}
\end{equation}
where $\zeta = F \sin(\theta_{\max}) / w $ . 	
In this work, we use a high-NA lens with $\mathrm{NA} = 0.8$,  $F = 1.95 \mathrm{mm}$, and set $\zeta = 8$ (or $w = 0.195 \mathrm{mm}$).

\section{\label{app:singleparticle} Single-particle Hamiltonian}
\begin{figure}[t!]
	\centering
	\includegraphics[width = 0.5\linewidth]{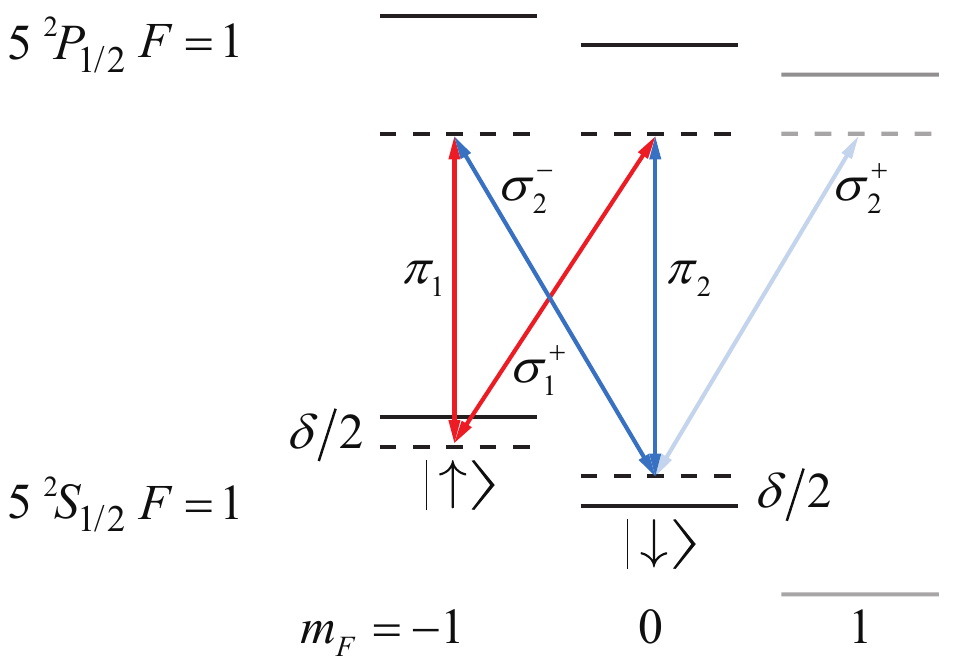}
	\caption{ 
		Double-$\Lambda$-type energy level configuration.
		Optical transitions allowed	by the selection rules involve the $\sigma_1^{+}$ and $\pi_2$ components or the $\pi_1$ and $\sigma_2^{-}$ components of the tightly focused VBs with optical frequencies $\omega_1$ (red) and $\omega_2$ (blue).  
	}
	\label{fig_EnergyLevel}
\end{figure}

We consider the double-$\Lambda$-type energy level configuration as shown in Fig.~\ref{fig:Scheme}~(c). 	
We choose the Zeeman sublevels $\left| \downarrow \right\rangle \equiv \left| 5^2S_{1/2}, F=1, m_F = 0 \right\rangle$ and $\left| \uparrow \right\rangle \equiv  \left| 5^2S_{1/2}, F=1, m_F=-1 \right\rangle$, and the excited states  $|5^2P_{1/2}, F=1, m_F = 0 \rangle$ and $|5^2P_{1/2}, F=1, m_F = -1 \rangle$. 
We plot the detailed double-$\Lambda$ scheme in \Fig{fig_EnergyLevel}. 
The states $\left| \uparrow \right\rangle$ and $\left| \downarrow \right\rangle$ are coupled via two-photon Raman processes, obeying the selection rules for the $\sigma_1^{+}$ and $\pi_2$ components, or the $\pi_1$ and $\sigma_2^{-}$ components of the tightly focused VBs.

The single-particle Hamiltonian is given by
\begin{equation}
	H_{0} =-\frac{\hbar^{2}\nabla^{2}}{2M}+\frac{1}{2}M\omega^{2}\rho^{2}+\frac{1}{2}M\omega_z^{2}z^{2}+h_{\mathrm{eff}},\label{eq:H0}
\end{equation}
where $h_{\mathrm{eff}}$ is the effective Hamiltonian for atoms in the ground-state
manifold during the Raman process, expressed as~\cite{Le-Kien2013, Goldman2014, Zhai2015, Zhai2021}
\begin{equation}
	h_{\mathrm{eff}}=u_s|\tilde{\vect{E}}|^2+\frac{\mu_Bg_F}\hbar\left(\mathbf{B}+\mathbf{B}_{\mathrm{eff}}\right)\cdot\mathbf{F}. \label{eq:heff2}
\end{equation}
The first term represents the scalar light shift, while the second term contains the effective magnetic field 
\begin{equation}
	\mathbf{B}_{\mathrm{eff}} = \frac{\mi u_v(\tilde{\vect{E}}^{*} \times \tilde{\vect{E}} )}{\mu_{B}g_{J}},
\end{equation} 
which is generated by the vector light shift. 
$u_s$ and $u_v$ are the scalar and vector polarizabilities, respectively~\cite{Le-Kien2013, Goldman2014, Zhai2015, Zhai2021}. 
The vector light shift induces coupling between different spin states. 
In our scheme, the total electric field experienced by the atoms is given by
\begin{equation}
	\tilde{\vect{E}} = \mathbf{E}_{1} \me^{\mi\omega_{1}t} +\mathbf{E}_{2} \me^{\mi\omega_{2}t},   \label{eq:Etotal}
\end{equation}
where $\mathbf{E}_{1} = (E_{x1},E_{y1},E_{z1})$ and $\mathbf{E}_{2} = (E_{x2},E_{y2},E_{z2})$. 
We find the potential induced by scalar light shift 
\begin{equation}
	V_{\mathrm{S}}(\mathbf{r})=u_{s}(\mathbf{E}_{1}^{*}\cdot \mathbf{E}_{1}+\mathbf{E}_{2}^{*}\cdot \mathbf{E}_{2}), \label{eq:appVr}
\end{equation}
and the effective potential $\mathbf{B}_{\mathrm{eff}}=(B_x, B_y, B_z)$ has components 
\begin{align}
	B_x & =\frac{\mi u_v}{\mu_{B}g_{J}}  (E_{y1}^{*}E_{z2} - E_{z1}^{*}E_{y2} ) \me^{-\mi (\omega_{1}-\omega_{2})t}  + \frac{\mi u_v}{\mu_{B}g_{J}} (E_{y2}^{*}E_{z1} -E_{z2}^{*}E_{y1}) \me^{\mi (\omega_{1}-\omega_{2})t},   \nn\\ 		
	B_y & =\frac{\mi u_v}{\mu_{B}g_{J}}  (E_{z1}^{*}E_{x2} - E_{x1}^{*}E_{z2}) \me^{-\mi (\omega_{1}-\omega_{2})t}  + \frac{\mi u_v}{\mu_{B}g_{J}} (E_{z2}^{*}E_{x1} - E_{x2}^{*}E_{z1}) \me^{\mi (\omega_{1}-\omega_{2})t} , \nn\\		
	B_z & = \frac{\mi u_v}{\mu_{B}g_{J}}(E_{x1}^{*}E_{y1}-E_{y1}^{*}E_{x1}+E_{x2}^{*}E_{y2}-E_{y2}^{*}E_{x2}).  \label{eq:BxBy}		
\end{align}	
Here we have neglected the terms corresponding to the forbidden optical transitions.

The effective Hamiltonian is reduced to~\cite{Goldman2014, Zhai2015, Zhai2021}
\begin{equation}
	h_{\mathrm{eff}}=V_{\mathrm{S}}(\mathbf{r})+\frac{\mu_{B}g_{F}}{2}(B_z\sigma_{z}+B_x\sigma_{x}+B_y\sigma_{y}).  \label{eq:heff3}
\end{equation}
After performing the rotating wave approximation and a unitary transformation with $U=e^{\mi (\omega_1-\omega_2)t\hat{\sigma}_z/2}$, the single-particle Hamiltonian becomes
\begin{align}
	H_0 &=-\frac{\hbar^{2}\nabla^{2}}{2M}+\frac{1}{2}M\omega^{2}\rho^{2}+\frac{1}{2}M\omega_z^{2}z^{2}+V_{\mathrm{S}}(\mathbf{r})+\frac{\delta}{2}\sigma_{z} +\begin{pmatrix}\Omega_{z}(\mathbf{r})&\Omega_{r}(\mathbf{r})\\\Omega_{r}^{*}(\mathbf{r})&-\Omega_{z}(\mathbf{r})\end{pmatrix},  \label{eq:H0final}
\end{align} 
where 
\begin{align}
	V_{\mathrm{S}}(\mathbf{r}) &= \Omega_s(\mathbf{\bar{E}}_{1}^{*}\cdot \mathbf{\bar{E}}_{1}+\mathbf{\bar{E}}_{2}^{*}\cdot \mathbf{\bar{E}}_{2}), \nonumber \nonumber \\
	\Omega_{z}(\mathbf{r}) &=\Omega_{0}\mi(\bar{E}_{x1}^{*}\bar{E}_{y1}-\bar{E}_{y1}^{*}\bar{E}_{x1}+\bar{E}_{x2}^{*}\bar{E}_{y2}-\bar{E}_{y2}^{*}\bar{E}_{x2}), \nonumber \\
	\Omega_{r}(\mathbf{r}) & =\Omega_{0}\big[(\bar{E}_{z2}^{*}\bar{E}_{x1}-\bar{E}_{x2}^{*}\bar{E}_{z1})+ \mi (\bar{E}_{y2}^{*}\bar{E}_{z1}-\bar{E}_{z2}^{*}\bar{E}_{y1}) \big],
\end{align}
and the coupling strengths are
\begin{align}
	\Omega_s &= u_s (I_1 + I_2 ), \\
	\Omega_{0} &= \frac{u_vg_{F} \sqrt{I_1 I_2}} {2g_{J}}. 
\end{align} 
Here $I_1$ and $I_2$ are the light intensities of the two VBs, and $\mathbf{\bar{E}}_{j}=\mathbf{E}_j/\sqrt{I_j}=(\bar{E}_{xj},\bar{E}_{yj},\bar{E}_{zj})$ is the normalized electric fields. 	
When interatomic interactions are weak, the motion of atoms can be approximated as a two-dimensional problem~\cite{Chen2020b, Zhang2019, Chen2018, Qu2015}. 
We focus on the physics near the focal plane at $z=0$, and the potentials approximately reduce to $V_{\mathrm{S}}(\vect{r}) = V_{\mathrm{S}}(\boldsymbol{\rho})$, $\Omega_{z}(\vect{r}) =  \Omega_{z}(\boldsymbol{\rho})$, and $\Omega_{r}(\vect{r}) = \Omega_{r}(\boldsymbol{\rho})$.

For the angular stripe phase in Fig.~\ref{fig:PhaseDiagram}, we adopt a wavelength of $\lambda = 797 \mathrm{nm}$. 
$\Omega_s$ is set to be finite and negative, inducing annular confinement via $V_{\mathrm{S}}(\boldsymbol{\rho})$. 
We find the ratio $ \Omega_0 / \Omega_s = -\frac{u_v\sqrt{I_1 I_2}} {8 u_s (I_1 + I_2)}$, and $u_v/u_s \approx 1.42$ for $\lambda = 797 \mathrm{nm}$~\cite{Le-Kien2013}. 
Given $\Omega_s$, $\Omega_0$ can be tuned by adjusting the intensities $I_1$ and $I_2$ of the VBs, making the angular stripe phase experimentally accessible.     
In Fig.~\ref{fig:Spintexture}, we adopt the tune-out wavelength $\lambda = 790 \mathrm{nm}$, where the scalar polarizability vanishes, so $\Omega_s \approx 0$. 
Here the coupling strength $\Omega_0$ is also tunable by adjusting two VBs' light intensities.

\section{\label{app:interactionBEC}Weakly interacting condensate}

We now consider the weakly interacting case.
The GP equation is given by
\begin{equation}
	\mi \hbar \frac{\partial}{\partial t} \Psi^{\mathrm{3D}} = (H_0+V_{\mathrm{I}}) \Psi^{3\mathrm{D}}, 
\end{equation}
where $V_{\mathrm{I}}$ represents the mean-field interaction: 
%\begin{equation}
%	V_{\mathrm{I}}(\mathbf{r}) = \begin{pmatrix} g_{\uparrow\uparrow}|\psi^{\mathrm{3D}}_{\uparrow}(\mathbf{r})|^2+g_{\uparrow\downarrow}|\psi^{\mathrm{3D}}_{\downarrow}(\mathbf{r})|^2  & 0 \\ 0 & g_{\uparrow\downarrow}|\psi^{\mathrm{3D}}_{\uparrow}(\mathbf{r})|^2+g_{\downarrow\downarrow}|\psi^{\mathrm{3D}}_{\downarrow}(\mathbf{r})|^2 \end{pmatrix}.  \label{eq:VI}
%\end{equation}	
\begin{equation}
	V_{\mathrm{I}}(\mathbf{r}) = \begin{pmatrix} g_{\uparrow\uparrow}|\psi^{\mathrm{3D}}_{\uparrow}|^2+g_{\uparrow\downarrow}|\psi^{\mathrm{3D}}_{\downarrow}|^2  & 0 \\ 0 & g_{\uparrow\downarrow}|\psi^{\mathrm{3D}}_{\uparrow}|^2+g_{\downarrow\downarrow}|\psi^{\mathrm{3D}}_{\downarrow}|^2 \end{pmatrix}.  \label{eq:VI}
\end{equation}	
Here $g_{\sigma\sigma^{\prime}}=4\pi\hbar^2Na_{\sigma\sigma^{\prime}}^s/M$ is the nonlinear interaction strength, $a_{\sigma\sigma^{\prime}}^s$ is the $s$-wave scattering length between two spins $\sigma = \uparrow, \downarrow$, and $N$ is total number of atoms.

In our scheme, we consider a pancake-shaped BEC with the trapping frequency $\omega_z \gg \omega$. 
The wave function is approximated as 
\begin{equation}
	\Psi^{\mathrm{3D}}=\begin{pmatrix}\psi _ \uparrow (\boldsymbol{\rho} )\\ \psi _ \downarrow (\boldsymbol{\rho} )\end{pmatrix} \psi ^{\mathrm{1D}}(z). \label{eq:phireduce}
\end{equation}
Inserting Eqs.~\eq{eq:H0final}, \eq{eq:VI} and \eq{eq:phireduce} into the GP equation, and integrating both sides with respect to $z$~\cite{Zhang2019}, we obtain
\begin{align}
	\mi\hbar \frac{\partial \psi _{\uparrow }}{\partial t} &=\left[ -\frac{\hbar
		^{2}\nabla _{\rho }^{2}}{2M}+\frac{1}{2}M\omega ^{2}{\rho }^{2}+V_{\mathrm{S}%
	}(\boldsymbol{\rho })+\frac{\delta }{2}+\Omega _{z}(\boldsymbol{\rho })%
	\right] \psi _{\uparrow }  \label{eq:GP1} +\eta \left( g_{\uparrow \uparrow }\left\vert \psi _{\uparrow }\right\vert
	^{2}+g_{\uparrow \downarrow }\left\vert \psi _{\downarrow }\right\vert
	^{2}\right) \psi _{\uparrow }+C\psi _{\uparrow }+\Omega _{r}(\boldsymbol{%
		\rho })\psi _{\downarrow }, \\
	\mi\hbar \frac{\partial \psi _{\downarrow }}{\partial t} &=\left[ -\frac{%
		\hbar ^{2}\nabla _{\rho }^{2}}{2M}+\frac{1}{2}M\omega ^{2}{\rho }^{2}+V_{%
		\mathrm{S}}(\boldsymbol{\rho })-\frac{\delta }{2}-\Omega _{z}(\boldsymbol{%
		\rho })\right] \psi _{\downarrow }  \label{eq:GP2} +\eta \left( g_{\uparrow \downarrow }\left\vert \psi _{\uparrow
	}\right\vert ^{2}+g_{\downarrow \downarrow }\left\vert \psi _{\downarrow
	}\right\vert ^{2}\right) \psi _{\downarrow }+C\psi _{\downarrow }+\Omega
	_{r}^{\ast }(\boldsymbol{\rho })\psi _{\uparrow }.
\end{align}
Here the constant $C$ and the dimensional reduction factor $\eta$ are defined as 
\begin{align}
	\eta &=\frac{\int_{z}\left\vert \psi^{\mathrm{1D}}(z)\right\vert^{4}dz}{%
		\int_{z}\left\vert \psi ^{\mathrm{1D}}(z)\right\vert ^{2}dz}, \qquad \\
	C &=\frac{\hbar ^{2}}{2M}\frac{\int_{z}\left\vert \frac{d\psi ^{\mathrm{1D}}(z)}{dz} 
		\right\vert ^{2}dz}{\int_{z}\left\vert \psi ^{\mathrm{1D}}(z)\right\vert ^{2}dz}+ 
	\frac{1}{2}M\omega _{z}^{2}\frac{\int_{z}z^{2}\left\vert \psi^{\mathrm{1D}}(z)\right\vert ^{2}dz}{\int_{z}\left\vert \psi ^{\mathrm{1D}}(z)\right\vert^{2}dz}.
\end{align}
Substituting $\psi_{\sigma}=\exp(-\mi C t/ \hbar) \tilde{\psi}_{\sigma}$, and replacing  $\tilde{\psi}_{\sigma} \rightarrow \psi_{\sigma}$, we obtain the reduced two-dimensional GP equation
\begin{align}
	\mi\hbar \frac{\partial \psi _{\uparrow }}{\partial t}& = \left[ -\frac{\hbar
		^{2}\nabla _{\rho}^{2}}{2M}+\frac{1}{2}M\omega ^{2}{\rho}^{2}+V_{\mathrm{S}%
	}(\boldsymbol{\rho})+\frac{\delta }{2}+\Omega _{z}(\boldsymbol{\rho} ) \right] \psi _{\uparrow } +\eta \left(
	g_{\uparrow \uparrow }\left\vert \psi _{\uparrow }\right\vert
	^{2}+g_{\uparrow \downarrow }\left\vert \psi _{\downarrow }\right\vert
	^{2}\right) \psi _{\uparrow } +\Omega _{r}(\boldsymbol{\rho} )\psi
	_{\downarrow }, \\
	\mi\hbar \frac{\partial \psi _{\downarrow }}{\partial t}& = \left[ -\frac{\hbar
		^{2}\nabla _{\rho}^{2}}{2M}+\frac{1}{2}M\omega ^{2}{\rho}^{2}+V_{\mathrm{S}%
	}(\boldsymbol{\rho})-\frac{\delta }{2}-\Omega _{z}(\boldsymbol{\rho} ) \right] \psi _{\downarrow }  +\eta \left(
	g_{\uparrow \downarrow }\left\vert \psi _{\uparrow }\right\vert
	^{2}+g_{\downarrow \downarrow }\left\vert \psi _{\downarrow }\right\vert
	^{2}\right) \psi _{\downarrow } +\Omega _{r}^{\ast }(\boldsymbol{\rho})\psi
	_{\uparrow }.
\end{align}
Moreover, the 1D wave function $\psi ^{\mathrm{1D}}(z)$ is given by form~\cite{Pethick2008}
\begin{equation}
	\psi ^{\mathrm{1D}}(z)= \frac{\mathrm{e}^{-z^2/2 a_{0z}^2}}{\pi^{1/4} a_{0z}^{1/2}},
\end{equation}
where $a_{0z} \equiv \sqrt{\hbar/M\omega_z}$ is the trap size along the $z$-axis. 
Therefore, the dimensional reduction factor is $\eta=1/ (\sqrt{2 \pi}a_{0z})$.

\section{\label{app:Potential} Potentials for varying VB parameters} 	
To analytically demonstrate the forms of $\Omega_{r}(\boldsymbol{\rho})$ and $\Omega_{z}(\boldsymbol{\rho})$, we substitute the electric fields (see~\ref{app:electricfields}) into \Eq{eq:VVB}. 
For simplicity, assuming both VBs have the same orientation angle $\beta_1 =\beta_2 = 0$, we obtain 
\begin{align}
	\Omega_{r}(\boldsymbol{\rho}) 
	& = f_{r}^{1}(\rho)\Theta_{r}^{1}(\alpha_1,\alpha_2)e^{\mi \bar{l}_{1}\varphi} + f_{r}^{2}(\rho)\Theta_{r}^{2}(\alpha_1,\alpha_2)e^{\mi \bar{l}_{2}\varphi}  + f_{r}^{3}(\rho)\Theta_{r}^{3}(\alpha_1,\alpha_2)e^{\mi \bar{l}_{3}\varphi} + f_{r}^{4}(\rho)\Theta_{r}^{4}(\alpha_1,\alpha_2)e^{\mi \bar{l}_{4}\varphi}, \\ 
	\Omega_{z}(\boldsymbol{\rho}) 
	& = f_{z}^{1}(\rho)\Theta_{z}^{1}(\alpha_1) + f_{z}^{2}(\rho)\Theta_{z}^{2}(\alpha_2) + f_{z}^{3}(\rho)\Theta_{z}^{3}(\alpha_1) + f_{z}^{4}(\rho)\Theta_{z}^{4}(\alpha_2) \nonumber \\
	& + f_{z}^{5}(\rho)\Theta_{z}^{5}(\alpha_1)e^{\mi \overline{l}_{5}\varphi} + f_{z}^{6}(\rho)\Theta_{z}^{6}(\alpha_1)e^{-\mi \overline{l}_{5}\varphi}  + f_{z}^{7}(\rho)\Theta_{z}^{7}(\alpha_2)e^{\mi \overline{l}_{6}\varphi} + f_{z}^{8}(\rho)\Theta_{z}^{8}(\alpha_2)e^{-\mi\overline{l}_{6}\varphi}, 
\end{align}
with spatial distributions
\begin{align}
	f_{r}^{1}(\rho)&=2\mi^{a_{1}+1}(-\mi)^{a_{2}}(\mathcal{I}_{a_{1}} \mathcal{I}'_{a_{2}+1} + \mathcal{I}'_{a_{1}+1} \mathcal{I}''_{a_{2}+2}), \nn\\ 
	f_r^2(\rho)&=2\mi^{a_1+1}(-\mi)^{b_2}(\mathcal{I}'_{a_1+1} \mathcal{I}_{b_2} - \mathcal{I}_{a_1} \mathcal{I}'_{b_2-1}), \nn\\ f_{r}^{3}(\rho)&=2\mi^{b_{1}+1}(-\mi)^{a_{2}}(\mathcal{I}''_{b_{1}-2}\mathcal{I}'_{a_{2}+1}-\mathcal{I}'_{b_{1}-1}\mathcal{I}''_{a_{2}+2}), \nn\\ f_{r}^{4}(\rho)&=-2\mi^{b_{1}+1}(-\mi)^{b_{2}}(\mathcal{I}'_{b_{1}-1}\mathcal{I}_{b_{2}}+\mathcal{I}''_{b_{1}-2}\mathcal{I}'_{b_{2}-1}), \nn\\ 
	f_{z}^{1}(\rho) &= -2(\mathcal{I}_{a_{1}}^{2} - \mathcal{I}_{a_{1}+2}''^{2}), \quad f_{z}^{2}(\rho)=-2(\mathcal{I}_{a_{2}}^{2}-\mathcal{I}_{a_{2}+2}''^{2}),  \nn\\
	f_{z}^{3}(\rho) &= 2(\mathcal{I}_{b_{1}}^{2} - \mathcal{I}_{b_{1}-2}''^{2}), \quad 
	f_{z}^{4}(\rho) = 2(\mathcal{I}_{b_{2}}^{2}-\mathcal{I}_{b_{2}-2}''^{2}), \nn\\
	f_{z}^{5}(\rho) &= (-\mi)^{b_{1}}\mi^{a_{1}}(\mathcal{I}''_{a_{1}+2}\mathcal{I}_{b_{1}} - \mathcal{I}_{a_{1}}\mathcal{I}''_{b_{1}-2}),  \nn\\
	f_{z}^{6}(\rho) &= (-\mi)^{a_{1}}\mi^{b_{1}}(\mathcal{I}''_{a_{1}+2}\mathcal{I}_{b_{1}} - \mathcal{I}_{a_{1}}\mathcal{I}''_{b_{1}-2}),  \nn\\
	f_{z}^{7}(\rho) &= (-\mi)^{b_2}\mi^{a_2}(\mathcal{I}''_{a_2+2}\mathcal{I}_{b_2} - \mathcal{I}_{a_2}\mathcal{I}''_{b_2-2}),  \nn\\
	f_{z}^{8}(\rho) &= (-\mi)^{a_{2}}\mi^{b_{2}}(\mathcal{I}''_{a_{2}+2}\mathcal{I}_{b_{2}} - \mathcal{I}_{a_{2}}\mathcal{I}''_{b_{2}-2}),  \nn\\
\end{align}
and angles
\begin{align}
	&\Theta_{r}^{1}(\alpha_1,\alpha_2)=\cos(\alpha_{1}/2)\cos(\alpha_{2}/2), \nn\\ 
	&\Theta_{r}^{2}(\alpha_1,\alpha_2)=\cos(\alpha_{1}/2)\sin(\alpha_{2}/2), \nn\\ &\Theta_{r}^{3}(\alpha_1,\alpha_2)=\sin(\alpha_{1}/2)\cos(\alpha_{2}/2), \nn\\ &\Theta_{r}^{4}(\alpha_1,\alpha_2)=\sin(\alpha_{1}/2)\sin(\alpha_{2}/2), \nn\\ 
	&\Theta_{z}^{1}(\alpha_1) = \cos^{2}(\alpha_1/2), \quad
	\Theta_{z}^{2}(\alpha_2) = \cos^{2}(\alpha_{2}/2),  \nn\\
	&\Theta_{z}^{3}(\alpha_1) = \sin^{2}(\alpha_1/2), \quad
	\Theta_{z}^{4}(\alpha_2) =\sin^{2}(\alpha_{2}/2), \nn\\
	&\Theta_{z}^{5}(\alpha_1) = \sin(\alpha_1), \quad
	\Theta_{z}^{6}(\alpha_1) = \sin(\alpha_1), \nn\\
	&\Theta_{z}^{7}(\alpha_2) = \sin(\alpha_2), \quad
	\Theta_{z}^{8}(\alpha_2) = \sin(\alpha_2). \nn
\end{align}
The $f$-functions with subscript $r$ only depend on the spatial variable $\rho$. 
Here $a_j=l_j - m_j$ and $b_j=l_j + m_j$ with $j=1,2$.
$\alpha_1$ and $\alpha_2$ denote the ellipticity angles on the high-order Poincaré sphere. 
We define the effective topological charges as:
\begin{align}
	\bar{l}_{1} &= l_1-l_2-m_1+m_2-1, \nn\\
	\bar{l}_{2} &= l_1-l_2-m_1-m_2+1, \nn\\
	\bar{l}_{3} &= l_1-l_2+m_1+m_2-3, \nn\\
	\bar{l}_{4} &= l_1-l_2+m_1-m_2-1, \nn\\
	\bar{l}_{5} &= 2-2m_1, \nn\\
	\bar{l}_6   &= 2-2m_2. 
\end{align}
By varying the VB parameters $(l_j, m_j, \alpha_j, \beta_j)$ with $j=1,2$, we can adjust the forms of $\Omega_{r}(\boldsymbol{\rho})$ and $\Omega_{z}(\boldsymbol{\rho})$, leading to different potentials for the atoms.
The specific forms of $\Omega_{r}(\boldsymbol{\rho})$ and $\Omega_{z}(\boldsymbol{\rho})$ for various VB parameters are presented in Table~\ref{tab:varyingVBParameters}.

\begin{table}[t!]
	%\centering
	\caption{\label{tab:varyingVBParameters}\centering $\Omega_{r}$ and $\Omega_{z}$ for different sets of VB parameters}
	\renewcommand{\arraystretch}{1.5} 
	\begin{tabular}{c|c|c} 
		\hline\hline
		\text{Scenario}  & $\Omega_{r}/\Omega_0$ & $\Omega_{z}/\Omega_0$ \\ \hline
		1. $\alpha_1=0,\alpha_2=0$ &$f_{r}^{1}e^{\mi \bar{l}_1\varphi}$&$f_{z}^{1}+f_{z}^{2}$  \\ \hline
		2. $\alpha_1=0,\alpha_2=\pi$ &$f_{r}^{2}e^{\mi \bar{l}_2\varphi}$& $f_{z}^{1}+f_{z}^{2}$ \\ \hline
		
		3. $\alpha_1=\pi,\alpha_2=0$ & $f_{r}^{3}e^{\mi \bar{l}_3\varphi}$ & $f_{z}^{3}+f_{z}^{4}$\\ \hline
		4. $\alpha_1=\pi,\alpha_2=\pi$ & $f_{r}^{4}e^{\mi \bar{l}_4\varphi}$ & $f_{z}^{3}+f_{z}^{4}$ \\ \hline
		
		5. $\alpha_1=0,\alpha_2=\forall$ & $[f_{r}^{1}\Theta_{r}^{1}+f_{r}^{2}\Theta_{r}^{2}]e^{\mi \bar{l}\varphi}$  & $f_{z}^{1}+f_{z}^{2}\Theta_{z}^{2}+f_{z}^{4}\Theta_{z}^{4}$  \\
		$Cond:m_2=1$&($\bar{l}= \bar{l}_{1}=\bar{l}_{2}$) &  $+f_{z}^{7}\Theta_{z}^{7}+f_{z}^{8}\Theta_{z}^{8}$ \\ \hline
		6. $\alpha_1=\pi,\alpha_2=\forall$ & $[f_{r}^{3}\Theta_{r}^{3}+f_{r}^{4}\Theta_{r}^{4}]e^{\mi \bar{l}\varphi}$ & $f_{z}^{3}+f_{z}^{2}\Theta_{z}^{2}+f_{z}^{4}\Theta_{z}^{4}$  \\
		$Cond:m_2=1$&($\bar{l}= \bar{l}_{3}=\bar{l}_{4}$) &  $+f_{z}^{7}\Theta_{z}^{7}+f_{z}^{8}\Theta_{z}^{8}$ \\ \hline
		
		7. $\alpha_1=\forall,\alpha_2=0$ & $[f_{r}^{1}\Theta_{r}^{1}+f_{r}^{3}\Theta_{r}^{3}]e^{\mi\bar{l}\varphi}$& $f_{z}^{2}+f_{z}^{1}\Theta_{z}^{1}+f_{z}^{3}\Theta_{z}^{3}$   \\
		$Cond:m_1=1$&($\bar{l}= \bar{l}_{1}=\bar{l}_{3}$) &  $+f_{z}^{5}\Theta_{z}^{5}+f_{z}^{6}\Theta_{z}^{6}$ \\ \hline
		8. $\alpha_1=\forall,\alpha_2=\pi$ & $[f_{r}^{2}\Theta_{r}^{2}+f_{r}^{4}\Theta_{r}^{4}]e^{\mi\bar{l}\varphi}$& $f_{z}^{4}+f_{z}^{1}\Theta_{z}^{1}+f_{z}^{3}\Theta_{z}^{3}$   \\
		$Cond:m_1=1$& ($\bar{l}= \bar{l}_{2}=\bar{l}_{4}$) &  $+f_{z}^{5}\Theta_{z}^{5}+f_{z}^{6}\Theta_{z}^{6}$ \\ \hline
		
		9. $\alpha_1 \neq 0,\pi;\alpha_2 \neq 0,\pi$ & \text{}   & \text{}     \\ 
		$Cond:m_1=m_2=1$ & $f(\rho)e^{\mi(2n-1)\varphi}$ &  0           \\  
		$l_1=n, l_2=-n$ \text{}   & $(n\in \mathbb{Z})$&      \\ 		\hline
		
		10. $\alpha_1 \neq 0,\pi;\alpha_2 \neq 0,\pi$ & \text{} & \text{} \\ 
		$Cond: m_1=m_2=1$ & $f(\rho)e^{\mi(n-1)\varphi}$ & $f_{z}(\rho)$         \\ 
		$l_1-l_2=n,l_1+l_2 \neq 0 $ & $(n\in \mathbb{Z})$ &  \text{} \\ \hline\hline			
	\end{tabular}
\end{table}

\textit{Scenarios 1-4}.  
$\Omega_r$ simplifies to a single term with a definite effective topological charge $\bar{l}$, representing the OAM transfer, while $\Omega_z$ reduces to a superposition of two spatially dependent terms. 

\textit{Scenarios 5-8}. 
If only one of $\alpha_1$ and $\alpha_2$ is $0$ or $\pi$, 
both $\Omega_r$ and $\Omega_z$ still exhibit a definite effective topological charge $\bar{l}$, but include additional spatially dependent terms in $\rho$. 

\textit{Scenario 9}. 
For the general cases, with $\alpha_1 \neq 0$ or $\pi$ and $\alpha_2 \neq 0$ or $\pi$, we consider the case $m_1 = m_2 = 1$ and $l_1 = - l_2 = n$, resulting in $\Omega_z = 0$ and $\Omega_{r}=\Omega_{0}f(\rho)e^{\mi (2n-1)\varphi}$, with $f(\rho)= \sum_{j' = 1}^4 f_{r}^{j'} \Theta_{r}^{j' }$. 
Here, $\Omega_{r}$ couples atomic spin and OAM, leading to an OAM transfer $\Delta l = (2n-1)$ for transition $\left|\uparrow\right\rangle$ $\rightarrow$ $\left|\downarrow\right\rangle$ and $\Delta l = -(2n-1)$ for the reverse.
This coupling resembles SOAMC~\cite{DeMarco2015, Sun2015, Qu2015, Hu2015, Chen2016, Chen2020a, Peng2022, Zhang2019, Chen2018}, with the OAM transfer tunable via VB parameters.

\textit{Scenario 10}. 
With $\alpha_1 \neq 0$ or $\pi$, $\alpha_2 \neq 0$ or $\pi$, $m_1 = m_2 = 1$, and $l_1 - l_2 = n$, the effective topological charge becomes $(n-1)$. 
$\Omega_z = \Omega_{0}f_z(\rho)$ introduces a spatially dependent Zeeman shift, where $f_z(\rho) = \sum_{j' = 1}^8 f_{z}^{j'} \Theta_{z}^{j' }$. 
Note that this feature is absent in LGB-induced scheme. 
Thus, VBs provide additional tunable degrees of freedom for the quantum control of ultracold atoms. 

\begin{figure}[t!]
	\centering
	\includegraphics[width = 0.7\linewidth]{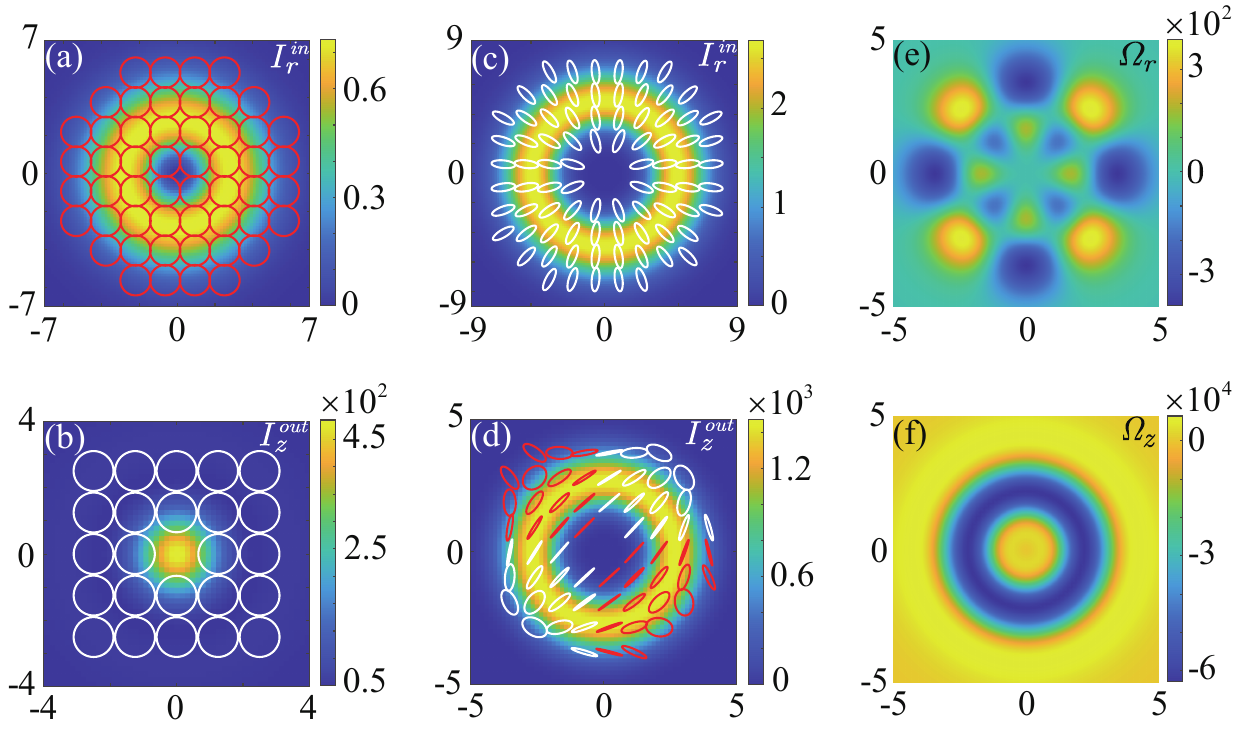}
	\caption{
		The polarization morphologies of incident VBs are shown for $j=1$ in (a) and $j=2$ in (c), and the tightly focused VBs on the focal plane for $j=1$ in (b) and $j=2$ in (d). 
		The background color indicates the light intensity $I_r^{\rm{in}}=|\boldsymbol{\mathcal{E}}|^2$ and $I_z^{\rm{out}}=|E_z|^2$.
		White and red represent left-handed elliptical and right-handed elliptical polarization states, respectively. 
		In (a) and (b), $m_1=2, l_1=-1, \alpha_1 = \pi$; in (c) and (d), $m_2=1, l_2=3, \alpha_2 = 0.3\pi$. 	
		The spatial distributions of $\Omega_r$ and $\Omega_z$ are shown in (e) and (f), respectively. 
		The spatial coordinates $x$ and $y$ are in units of trap size $a_0$. 
	}
	\label{fig_Polarization}
\end{figure}

In \Fig{fig_Polarization}, we show the polarization and spatial distributions for specific VB parameters. 
In \Fig{fig_Polarization} (a), for $\alpha= 0$ or $\pi$, the polarization distribution is homogeneous, and the VB corresponds to regular left- or right-circularly polarized light~\cite{Rosales2018}. 
After passing through the tightly focusing system, the polarization reverses, as shown in \Fig{fig_Polarization} (b).

For $\alpha \neq 0$ or $\pi$, the beam exhibits a superposition of left- and right-circularly polarized components, resulting in a VB with nonuniform polarization distribution, as shown in \Fig{fig_Polarization} (c).  
On the focal plane, this polarization distribution becomes more complex, as depicted in \Fig{fig_Polarization} (d). 
Additionally, the tight-focusing system also reduces spot size and increases intensity. 
Using these tightly focused VBs to couple atomic pseudospin levels, we obtain the spatial distributions of $\Omega_r$ and $\Omega_z$ shown in Fig.~\ref{fig_Polarization} (e) and (f).

\section{\label{app:Skyrmion} Giant skyrmions and ground-state density profiles} 
\begin{figure*}[t!]
	\centering
	\includegraphics[width = 0.7\linewidth]{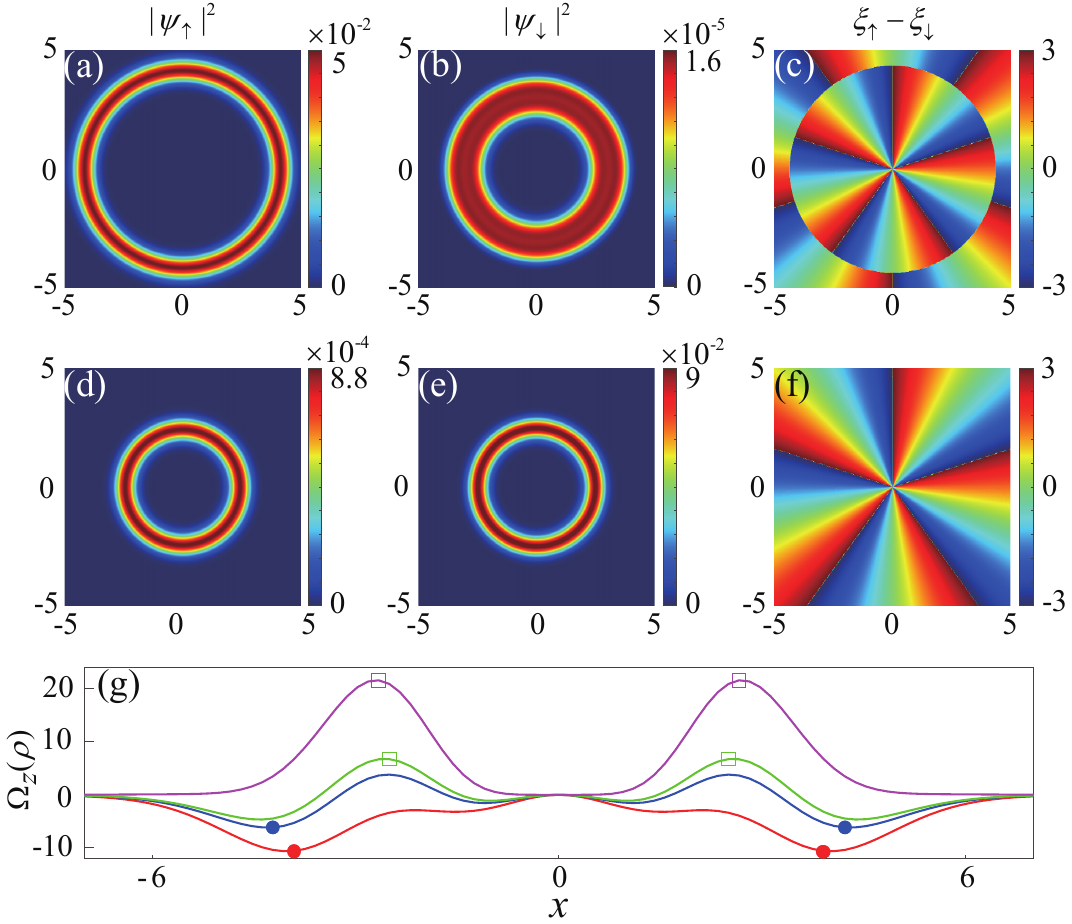}
	\caption{The ground state density profile $|\psi_\sigma|^2 a_0^2 /N$ and relative phase $\xi_\uparrow - \xi_\downarrow$ within the spatial range $x,y \in [-5 a_0, 5 a_0]$ for different ellipticity angles: (a)-(c) $\alpha_{1}=\alpha_{2}=0.37\pi$, (d)-(f) $\alpha_{1}=\alpha_{2}=0.45\pi$. 
		(g) The spatial distribution of $\Omega_{z}$ along the $x$ axis at $y = 0$.  
		Red, blue, green, and magenta correspond to $\alpha_{1}=\alpha_{2}=0, 0.37\pi, 0.45\pi$, and $\pi$, respectively.  
		Circle (square) labels the locations of maximum population of spin-up (spin-down) component. 
		We set the other parameters $\beta_1=\beta_2=0$, $m_1=1$, $l_1=2$, $m_2=1$, $l_2=-4$, $\delta=0$, $\Omega_{0} = 2.58 \hbar \omega$, $\Omega_s = 0$. 
		The coordinates $x$ and $y$ are in units of trap size $a_0$. }
	\label{fig:ST_density}
\end{figure*}

We analyze the spin textures in detail. 
Distinct from the SOAMC induced by LGBs~\cite{DeMarco2015, Sun2015, Qu2015, Hu2015, Chen2016, Chen2020a, Peng2022, Zhang2019, Chen2018}, our scheme introduces a nonzero $\Omega_z$, resulting in a spatially dependent Zeeman shift term $\Omega_z\sigma_z$, providing an additional tunable degree of freedom.
The spatial profile of $\Omega_z$ resembles a ring-shaped potential, as illustrated in \Fig{fig:ST_density} (g). 
The spin-up and spin-down components experience opposing external potentials, leading to a spin-polarized ground state. 
As $\alpha_1$ and $\alpha_2$ varies from 0 to $\pi$, the population peak locations shift, altering the density distribution of each spin component, with the positions of maximum population of spin-up or spin-down component labeled in \Fig{fig:ST_density} (g).
For $\alpha_1 = \alpha_2 = 0$ or $\pi$, the ground state becomes fully polarized with $\langle \sigma_z \rangle = \pm 1$, representing a topologically trivial structure. 
However, for intermediate values with $0 < |\langle \sigma_z \rangle| < 1$, the spin imbalance between the two spin states decreases. 
Specifically, for $\alpha_1 = \alpha_2 = 0.37\pi$ and $0.45\pi$, we illustrate the density distributions and relative phases in \Fig{fig:ST_density} (a)-(f). 
The relative phases in \Fig{fig:ST_density} (c) and (f) show that the ground state exhibits a stable multiply quantized vortex with its quantized circulation to be $2\pi\times 5 $. 

The corresponding spin textures, as shown in Fig.~\ref{fig:Spintexture}~(c) and (d), reveal topologically nontrivial giant skyrmion structures. 
To calculate the topological charge of skyrmion, we define a normalized complex-valued spinor $\mathbf{\chi} = [\chi_{\uparrow},\chi_{\downarrow}]^{T} = \left[|\chi_{\uparrow}| \mathrm{e}^{\mi\xi_{\uparrow}},|\chi_{\downarrow}| \mathrm{e}^{\mi \xi_{\downarrow}}\right]^{T}$, satisfying $|\chi_{\uparrow}|^{2}+|\chi_{\downarrow}|^{2}=1$. 
The wave functions can be expressed as $\psi_{\sigma}=\sqrt{\rho_{T}}\chi_{\sigma}$, with total density $\rho_{T}=|\psi_{\uparrow}|^{2} + |\psi_{\downarrow}|^{2}$. 
The pseudospin density is defined as $\mathbf{S}=\mathbf{\chi}^\dagger \vec{\sigma}_{\rm P} \mathbf{\chi}$ with $\vec{\sigma}_{\rm P}$ the Pauli matrix~\cite{Wang2017,Jin2013,Yang2008,Su2024}. The components of $\mathbf{S}$ are expressed as 
\begin{equation}
	\begin{aligned}
		&S_{x} =2\left|\chi_{\uparrow}\right|\left|\chi_{\downarrow}\right|\cos(\xi_{\uparrow}-\xi_{\downarrow}), \\
		&S_{y} =-2\left|\chi_{\uparrow}\right|\left|\chi_{\downarrow}\right|\sin(\xi_{\uparrow}-\xi_{\downarrow}), \\
		&S_{z} =|\chi_{\uparrow}|^{2}-|\chi_{\downarrow}|^{2} , 
	\end{aligned}
\end{equation}
and $|\mathbf{S}|^{2}=S_{x}^{2}+S_{y}^{2}+S_{z}^{2}=1$. 
In polar coordinates, the spin vector $\mathbf{S}$ can be written as 
\begin{equation}
	\vect{S}=( \sqrt{1-S_{z}^{2}}\cos(\Delta\kappa\varphi),\sqrt{1-S_{z}^{2}}\sin(\Delta\kappa\varphi),S_{z} ),
\end{equation}
where the phases of two spin components can be approximately written as $\xi_{\sigma}=\kappa_{\sigma} \varphi$, with $\kappa_{\sigma} = 0,1,2,...$ denoting the quantum number of the circulation of the spin component at radius $\rho$, and $\Delta\kappa=\kappa_{\uparrow}-\kappa_{\downarrow}$.
For $\alpha_{1}=\alpha_{2}=0.37\pi$, $\kappa_{\uparrow}=5$, and $\kappa_{\downarrow}=0$ at the radii of both the inner and outer annular giant skyrmions. 
The corresponding topological charge density is written as $q(\boldsymbol{\rho})=\frac{\Delta\kappa}{4\pi \rho}\frac{dS_z(\rho)}{d\rho}$.
Then we find the topological charges of giant skyrmions $Q_{\mathrm{in}} =\int \frac{\Delta\kappa}{4\pi \rho}\frac{dS_z(\rho)}{d\rho} \mathrm{d}\boldsymbol{\rho}=(\kappa_{\uparrow}-\kappa_{\downarrow})_{\mathrm{in}}=5$ and $Q_{\mathrm{out}} =-(\kappa_{\uparrow}-\kappa_{\downarrow})_{\mathrm{out}} =-5$.
Increasing $\alpha_1$ and $\alpha_2$ to $\alpha_1=\alpha_2=0.45\pi$, 
we find a single giant skyrmion with topological charge $Q=-5$. 
Additionally, we perform calculations in the absence of $\Omega_z$, and find topologically trivial spin textures. 
This shows that the rich topological structures arise from the interplay of $\Omega_z$ and $\Omega_r$, demonstrating the high tunability of VB-induced gauge fields in exploring topological quantum phenomena.

%\bibliographystyle{elsarticle-num}
%\bibliographystyle{cas-model2-names}
%\bibliography{TFVB_reference}

\end{document}